\documentclass[conference]{IEEEtran}
\IEEEoverridecommandlockouts
\usepackage{cite}
\usepackage{amsmath,amssymb,amsfonts}
\usepackage{graphicx}
\usepackage{textcomp}
\usepackage[table,xcdraw]{xcolor}
\usepackage{pifont}
\newcommand{\xmark}{\ding{55}}%
\usepackage{float}
\usepackage{algorithm}
\usepackage{algpseudocode}

\def\BibTeX{{\rm B\kern-.05em{\sc i\kern-.025em b}\kern-.08em
    T\kern-.1667em\lower.7ex\hbox{E}\kern-.125emX}}
\begin{document}

\title{5G-SRNG: 5G Spectrogram-based Random Number Generation for Devices with Low Entropy Sources}

\author{
\IEEEauthorblockN{Ferhat Ozgur Catak}
\IEEEauthorblockA{\textit{Electrical Engineering and Computer Science} \\
\textit{University of Stavanger}\\
Stavanger, Norway \\
f.ozgur.catak@uis.no}
\and
\IEEEauthorblockN{Evren Catak}
\IEEEauthorblockA{\textit{Independent researcher} \\
Stavanger, Norway \\
evren.catak@ieee.org}
\and
\IEEEauthorblockN{Ogerta Elezaj}
\IEEEauthorblockA{\textit{School of Computing and Digital Technology} \\
\textit{Birmingham City University}\\
Birmingham, United Kingdom \\
ogerta.elezaj@bcu.ac.uk}

}

\maketitle

\begin{abstract}
Random number generation (RNG) is a crucial element in security protocols, and its performance and reliability are critical for the safety and integrity of digital systems. This is especially true in 5G networks with many devices with low entropy sources. This paper proposes 5G-SRNG, an end-to-end random number generation solution for devices with low entropy sources in 5G networks. Compared to traditional RNG methods, the 5G-SRNG relies on hardware or software random number generators, using 5G spectral information, such as from spectrum-sensing or a spectrum-aware feedback mechanism, as a source of entropy. The proposed algorithm is experimentally verified, and its performance is analysed by simulating a realistic 5G network environment. Results show that 5G-SRNG outperforms existing RNG in all aspects, including randomness, partial correlation and power, making it suitable for 5G network deployments.
\end{abstract}


\section{Introduction}
With the increasing demand for higher data rates, lower latency, and more extensive coverage, 5G networks are increasingly becoming the standard for wireless communications \cite{9527756}. However, one of the main challenges for 5G networks is security, as 5G networks are expected to be deployed in large, heterogeneous and distributed environments. To address this challenge, secure communication protocols are needed to provide secure communication between nodes and protect against malicious attacks.

Random number generation (RNG) is crucial for cryptographic systems and communication security, as it is used in a wide range of applications, from digital signatures and encryption protocols to password generation, game development and data transfer. However, in 5G networks, the generated random numbers are unreliable, due to the many devices in 5G networks with low entropy sources, which may be susceptible to attack or malfunction. The quality of random numbers plays a significant role in the security of cryptographic protocols, ranging from Elliptic Curve Diffie-Hellman (ECDH) key exchange algorithms to Rijndael encryption algorithms, which are used in many 5G networks. Many cryptographic protocols have been broken due to low-quality random numbers. Hence, it is essential to develop an RNG solution to provide more secure protocols in 5G networks.

 As many cryptographic protocols depend on the generation of random numbers, and random numbers are the input for generating the keys, the quality of random numbers plays a significant role in the security of cryptographic protocols\cite{8600398}. In a key exchange protocol like Diffie-Hellman \cite{1055638}, the private key should be based on high-quality random numbers. Many key exchange protocols have been broken due to low-quality random numbers. 

RNG is a device that generates random numbers. Many RNGs have been proposed in the literature \cite{9181042,9375939,9043369,9190421}. RNG can be categorised as either true or pseudorandom,depending on how they are realized, with each type having different properties. A true RNG, known as a physical RNG,  is a device that generates random numbers from a physical process, such as radioactive decay or noise. A pseudo RNG is a device that generates random numbers that appear random but are in fact deterministic. They are often utilized in computer programs and they are cryptographically secure if it is computationally infeasible to predict the next number in the sequence. 

IoT devices' entropy sources are limited due to their limited hardware resources. There is great potential to use the existing 5G signals to generate entropy for IoT devices. IoT devices can capture the 5G signals using an antenna and use this signial to generate entropy. 

This paper proposes 5G-SRNG, an end-to-end RNG method, with a diffrent approach of traditional RNGs, for devices with low entropy sources in 5G networks,providing more secure communication. Instead of relying on hardware or software RNGs, 5G-SRNG uses 5G spectral information, such as from spectrum-sensing or a spectrum-aware feedback mechanism, as a source of entropy. The 5G-SRNG method uses a novel approach to generate random numbers using the 5G spectrograms produced by the spectrum-sensing process. The proposed solution it is experimentally verified using NIST tests, in a simulated realistic 5G network environment. The main contributions of this paper are summarized as follows:

\begin{itemize}
\item It proposes 5G-SRNG, an end-to-end RNG solution for 5G networks with low entropy sources.
\item It analyses the performance of 5G-SRNG in a realistic 5G network environment.
\item It discuss the advantages of 5G-SRNG over existing RNG solutions.
\item The experimental results shows the seperior performance of 5G-SRNG.
\end{itemize}

The rest of this paper is organized as follows. Section \ref{sec:relworks} presents the related work. Section \ref{sec:preliminaries} describes the preliminary work. Section \ref{sec:system_model} describes the 5G-SRNG. Section \ref{sec:experiments} shows the experimental results. Finally, the conclusion and future work are given in the Section \ref{sec:conclusion}.

\section{Related Work}\label{sec:relworks}
Many RNGs are based on physical processes, such as thermal noise, radioactive decay, and other physical phenomena \cite{ruhkin2001testing,9181042,9375939}. True RNGs have good entropy and can generate high-quality random numbers. The physical process limits the true RNGs they are based on. The main disadvantages of the true RNGs are being slowly and expensivly making them unusable for most computing applications. Ont he other side, pseudo-RNGs are more efficient and less costly than true RNGs,and for these reasons they are widely used in cryptography to generate keys for exchange protocols \cite{4568378}. A typical key exchange protocol is the Diffie-Hellman protocol. Entropy, as a measure of unpredictability, it is used to of an RNG is an important property. The entropy of an RNG determines the quality of random numbers. The entropy of an RNG is a measure of the unpredictability of the random numbers. There are many methods to measure the entropy of an RNG. The most common form is NIST Statistical Test Suite. Entropy sources are used to improve the quality of random numbers. Entropy sources are used to increase the unpredictability of random numbers.  


There are many works on RNG in the literature \cite{9181042,9375939,9043369,9190421}. Still, there is no work specifically on RNG for devices with low entropy sources in 5G networks. 

\section{Preliminaries}\label{sec:preliminaries}

\subsection{5G and beyond in IoT and Security Requirements}

5G networks can provide a wide range of services and applications in the IoT domain. These services and applications include connected cars, smart homes, and industrial automation. 5G has several advantages over its predecessor technologies, such as higher data rates, lower latency, distributed architectures, and higher capacity. 5G networks are being deployed to support the growth of IoT and other technologies.

The implementation of robust security measures is crucial for ensuring the integrity and confidentiality of 5G networks, as they are equally susceptible to various forms of cyber threats and attacks. There are three principal security concerns that are of particular relevance to 5G networks: protocol stack vulnerabilities exploitation by attackers, compromise of messages and authentication mechanisms, and physical layer attacks. To effectively mitigate these risks, the use of authentication and cryptography protocols are imperative. These protocols rely on the generation of high-quality, random numbers to produce cryptographic keys, which must be generated through reliable and robust random number generators (RNGs).

\subsection{Random Numbers in Cryptography}
Cryptography is the science of secure communication. Cryptography is used to protect information from unauthorized access. Cryptography is used in many applications, such as electronic commerce, email, and secure communications. Cryptography is used to secure communications between two parties. An cryptographic protocol is a triple of algorithms $(\mathsf{KeyGen}, \mathsf{Enc}, \mathsf{Dec})$:

\begin{itemize}
    \item The key generation algorithm  $\mathsf{KeyGen}$ is used to generate keys. The key generation algorithm uses an entropy source to produce high-quality random numbers. The entropy source is used to generate keys. The key generation algorithm outputs the public key, $key_{pub}$, and the private key, $key_{priv}$. The entropy source is used to generate the public and private keys.
    \item The encryption algorithm $\mathsf{Enc}$ is used to encrypt a message using the public key. The encryption algorithm has two inputs: a public key $key_{pub}$ and a message, $M$. 
    \item The decryption algorithm $\mathsf{Dec}$ is used to decrypt the message using the private key. The decryption algorithm has two inputs: a private key $key_{priv}$ and a message, $M^{\prime}$. 
\end{itemize}

The encryption and decryption algorithms $\mathsf{Enc}$ and $\mathsf{Dec}$ are invertible. That is, if $M^{\prime} = \mathsf{Enc}(M, K_{pub})$, it follows that  $M = \mathsf{Dec}(M^{\prime}, K_{priv})$. The recipient uses the private key to decrypt the message, and the attacker uses the public key to decrypt the message. The security of a cryptographic protocol depends on the encryption and decryption algorithms, the secret key. 


The paramount attribute of a RNG is the quality of randomness it exhibits. The degree of randomness in an RNG is a metric that indicates the unpredictability of the generated random numbers. Ideal random numbers must be uniformly distributed, meaning that each numerical value is generated with an equal likelihood. Additionally, it is essential that the random numbers produced exhibit low correlation with consecutive values.

The security of a cryptographic protocol is dependent upon the quality of the RNG utilized in the generation of the secret key. RNGs are responsible for generating random numbers, which can either be true or pseudorandom. True RNGs, also referred to as physical RNGs, generate random numbers through physical processes such as radioactive decay or thermal noise. On the other hand, pseudorandom RNGs, also referred to as deterministic RNGs, generate random numbers through a deterministic process. In cryptography, pseudorandom RNGs are commonly employed for key generation, and it is imperative that they possess cryptographic security. A pseudorandom RNG is deemed cryptographically secure if it is computationally intractable to predict the subsequent number in the sequence.

In the context of operating systems, RNGs are utilized to generate seed values. In the Linux operating system, the RNG "/dev/urandom" is employed for this purpose. The seed values generated by an entropy source, such as mouse movements or keyboard events, are used to initialize the "/dev/urandom" RNG. This RNG produces a non-blocking stream of pseudorandom numbers. However, in the context of Internet of Things (IoT) devices, the entropy sources available may be limited, making it infeasible to use "/dev/urandom" for cryptographic protocols aimed at protecting the communication between IoT devices. In such scenarios, alternative RNGs or entropy sources may need to be considered to ensure the cryptographic security of the system.

The traditional RNGs suffer from two problems \cite{9631213}: 
\begin{itemize}
    \item \textbf{Low Entropy}: Traditional RNGs are not suitable for cryptographic applications. They have low entropy, which means they are not secure.
    \item \textbf{Predictability}: Traditional RNGs are predictable. The output of the RNG can be predicted if the internal state of the RNG is known.
\end{itemize}

A significant number of cryptographic protocols have been compromised due to the usage of low-quality random numbers. To mitigate such risks, it is imperative that pseudorandom number generators (PRNGs) utilized in cryptography exhibit cryptographic security. In numerous cryptographic protocols, the security of the system is directly tied to the quality of the random numbers generated. Hence, prior to their usage in cryptographic protocols, it is essential to thoroughly evaluate and verify the quality of the random numbers generated by the PRNGs.

\subsection{How to test randomness}
Randomness testing is an evaluation process aimed at determining the quality of random numbers. This testing is performed to ascertain if the random numbers generated by a RNG are of high quality. The quality of the random numbers can be assessed through the evaluation of the RNG through randomness tests. The higher the quality of the random numbers generated, the greater the security of the cryptographic system.

A variety of techniques exist for assessing the quality of random numbers, with the NIST Statistical Test Suite \cite{10.5555/2206233} being the most widely employed method. This open-source test suite is designed to evaluate the randomness of RNGs by conducting statistical tests on the generated numbers.

The NIST Statistical Test Suite is a widely adopted tool for evaluating the quality of random numbers. It comprises of 15 statistical tests, which are grouped into two categories. These tests are used to assess the randomness of the numbers generated by RNGs.

\begin{itemize}
    \item The initial tests (Frequency, Block Frequency, Cumulative Sums, Runs, and Longest Runs);
    \item The remaining tests (Binary Matrix Rank, Discrete Fourier Transform, Non-overlapping Template Matching, Overlapping Template Matching, Maurer's "Universal Statistical", Linear Complexity, Serial, Approximate Entropy, Random Excursions, and Random Excursions Variant).
\end{itemize}

The NIST Statistical Test Suite's 15 statistical tests are designed to assess the presence of non-random patterns in a set of random numbers generated by RNGs. These tests thoroughly evaluate the randomness of the numbers by checking for a comprehensive range of potential non-randomness characteristics.
\begin{itemize}
    \item \textbf{The frequency test} checks for a uniform distribution of the numbers.
    \item \textbf{The block frequency test} checks for clusters of consecutive numbers. 
    \item \textbf{The cumulative sums test} checks for trends in the set of numbers. 
    \item \textbf{The runs test} checks for runs of consecutive numbers.
    \item \textbf{The longest runs} test checks for runs of consecutive numbers. 
    \item \textbf{The binary matrix rank test} checks for patterns in a binary matrix.
    \item \textbf{The discrete Fourier test} checks for patterns in the frequency domain.
    \item \textbf{The non-overlapping template matching test} checks for patterns of a particular length.
    \item \textbf{The overlapping template matching test} checks for patterns of a particular length.
    \item \textbf{The Maurer's universal statistical test} checks for non-randomness in the statistical properties of the sequence. 
    \item \textbf{The linear complexity test} checks for linear dependence in a sequence.
    \item \textbf{The serial test} checks for non-randomness in the shift of a periodic autocorrelation.
    \item \textbf{The approximate entropy test} checks for non-randomness in the complexity of a sequence.
    \item \textbf{The random excursions test} checks for non-randomness in the statistical properties of the sequence. 
    \item \textbf{The random excursions variant test} checks for non-randomness in the statistical properties of the sequence.
\end{itemize}

It is imperative to note that an RNG that passes the NIST Statistical Test Suite does not guarantee that it is a Cryptographically Secure pseudo RNG (CSPRNG). The tests are designed to detect patterns and deviations from randomness, but there may still be potential weaknesses that are not covered by the tests. Hence, it is important to continuously monitor and evaluate the quality of random numbers generated by RNGs in real-world implementations. This is crucial for ensuring the security of cryptographic protocols and the confidentiality of sensitive information transmitted over networks.

\section{System Model}\label{sec:system_model}
\begin{figure*}[!htbp]
    \centering
    \includegraphics[width=0.89\linewidth]{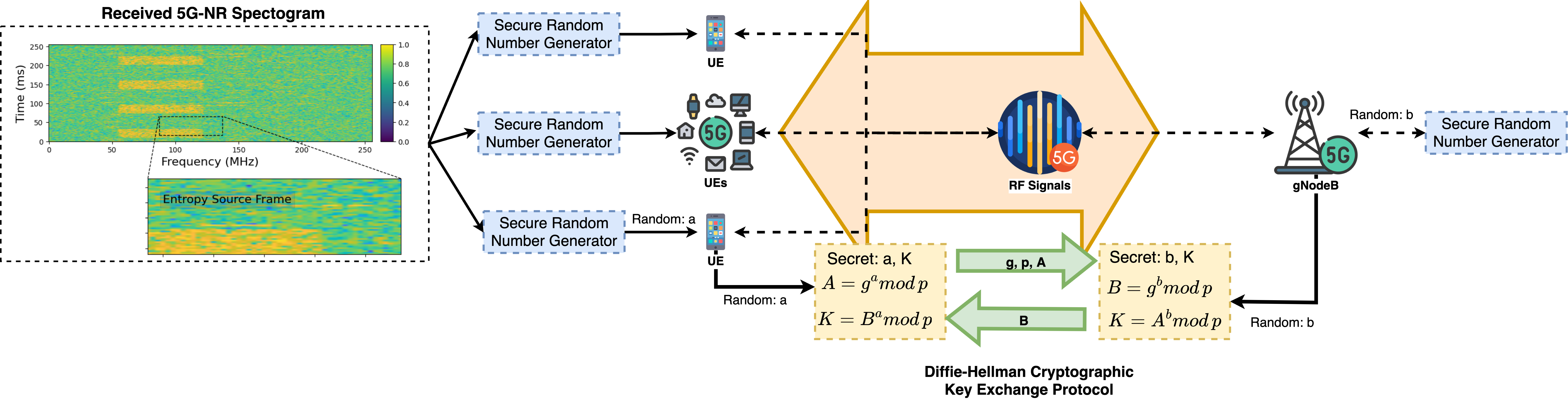}
    \caption{System model}
    \label{fig:sys_model}
\end{figure*}

The proposed algorithm represents a novel approach to generating random numbers by utilizing the entropy present in pilot signals. The process involves the random selection of a spectrum, the computation of its magnitude, and the extraction of entropy from the resulting spectrogram to generate a sequence of random numbers. The system model for the proposed approach is depicted in Figure \ref{fig:sys_model}. Using pilot signals as a source of entropy enhances the randomness and unpredictability of the generated random numbers, making them suitable for use in cryptographic protocols.

The next step is to extract the entropy from the spectrogram. The entropy is calculated by measuring the unpredictability of the spectrum magnitude. The entropy is used to generate a random sequence of numbers. The generated random numbers should pass the randomness tests to ensure their quality and suitability in cryptographic protocols. The proposed algorithm can provide a secure and efficient method for generating random numbers in IoT devices, as the pilot signals are a readily available resource in wireless communication systems. The proposed approach can improve the security of cryptographic protocols by providing high-quality random numbers to generate secret keys.

\subsection{Algorithm for Proposed 5G-SRNG}

\begin{figure}
    \centering
    \includegraphics[width=0.85\linewidth]{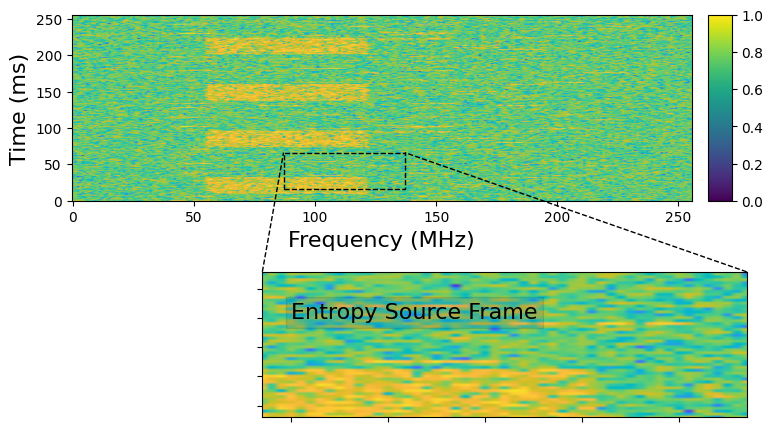}
    \caption{Example entropy source generation procedure from 5G NR signals.}
    \label{fig:my_label}
\end{figure}

Algorithm~\ref{rotation} shows the pseudocode of the proposed 5G-SRNG. We defined $m$ as the size of the 5G Spectrogram's ($\mathcal{D}$) rows,  $n$ as the size of the columns and $frame$ as a smaller $\mathcal{D}$. We defined $x_{start}, y_{start}$ as two starting points to select the random part (i.e. convolution) of $\mathcal{D}$. We defined $x_{end}, y_{end}$ as two ending points of the selected $frame$. $x_{start}$,  $x_{end}$, $y_{start}$ and $y_{end}$ are random entries of $\mathcal{D}$. We defined $frame$ as a smaller fixed-size $\mathcal{D}$ from 5G Spectrogram. $frame$ contains $c \times k$ elements (i.e. pixels) from $\mathcal{D}$ with $c \ge 1$ and $k \ge 1$. $frame[i, j]$ is the substantial single element (i.e. pixel) of $frame$. $seed$ is the main contribution of 5G-SRNG. We defined iterations ($t$) as the loop through the selected $frame$'s pixels. {\bfseries XOR} is a simple operator. {\bfseries shifting} is a simple operator on bits. $seed$ contains the total value of the XORed $frame$'s pixels. $c$ and $k$ are the variables we can control to generate random numbers by using a loop into $\mathcal{D}$. $c$ and $k$ can be different for 5G Spectrogram with different sizes.

\begin{algorithm}[!htbp]
    \label{rotation}
  \caption{Proposed 5G-SRNG}
  \begin{algorithmic}[1]
  \Require 5G Spectrogram: $\mathcal{D} \in \mathbb{R}^{m \times n}$, frame size: $c, k$
  \State $x_{start}, y_{start}  \gets random(0,m), random(0,n) $
  \State $x_{end} = min(x_{start} + c, m) $
  \State $x_{end} = min(y_{start} + k, n) $
  \State $frame \gets  \mathcal{D}[x_{start}:x_{end}, y_{start}:y_{end}].flatten()$
  \State $seed \gets 0$ \Comment{initialize $seed$ with 32-bit float representation of 0}
  \For {$t \in frame$}
    \State $seed \gets seed \mathbin{\oplus} t$ \Comment{XOR operator}
    \State $seed \gets seed \mathbin{\oplus} seed<<13$ \Comment{shift-left the previous $seed$ value 13 bits, then perform XOR operator}
    \State $seed \gets seed \mathbin{\oplus} seed>>17$ \Comment{shift-right the previous $seed$ value 17 bits, then perform XOR operator}
    \State $seed \gets seed \mathbin{\oplus} seed<<5$ \Comment{shift-left the previous $seed$ value 5 bits, then perform XOR operator}
  \EndFor
  \State \Return $seed$
    \end{algorithmic}
\end{algorithm}

The algorithm in the pseudocode can be used to generate random numbers from the 5G spectrogram. It works by measuring the 5G spectrogram through spectrum sensing and then randomly selecting a frame (convolution) within the spectrogram. The values of the pixels in the specified frame are used as the seed for the random number generation process. The seed is then XORed with its shifted values and converted into a 32-bit float representation of a random number. The algorithm shows how a 5G scale entropy source can be generated and perceived. The strength of the frequencies at a particular time is represented by each point in the two-dimensional array of the 5G spectrogram. The algorithm can generate random numbers from the spectrum magnitude by randomly selecting frames from the spectrogram.

Additionally, the XOR operation can be considered a non-linear operation that can produce pseudo-random numbers from deterministic inputs  \cite{masood2022lightweight}. The XOR operation is used multiple times in the algorithm, each time with different shifted values of the seed. This leads to multiple layers of chaotic operations and produces a highly unpredictable and random output. The unpredictability of the XOR operation is an essential factor that ensures the randomness of the generated numbers in 5G-SRNG.

Shift-left and shift-right operations respectively expand or contract the set of possible outcomes of the $seed$ value, as they delay or replicate the bits in the $seed$ while maintaining their magnitude. The multiplication or division of the $seed$ value by 2 results in an exponential expansion or contraction of the range of possible outcomes by a factor of $2^1$ or $2^{-1}$, respectively. The XOR and shift operations applied to the $seed$ value result in a strengthened and improved quality of the random number generated compared to conventional random number generators.

\subsection{Dataset Description}


The dataset used in this study was synthesized using the MATLAB 5G Toolbox. The frames, each of duration 40 milliseconds, were randomly shifted in the frequency domain. It was assumed that the 5G signals were within the specified frequency band range, and the network performance was evaluated based on the varying random bands. The sampling rate of 61.44 MHz was deemed sufficient to process 5G signals effectively. To generate the respective $256 \times 256$ RGB spectrogram images, the complex baseband signals were transformed using a Fast Fourier Transform (FFT) with a length of 4096.

Parameters of 5G signal generation are indicated in Table \ref{tab:datasetParameter}. In the table, SCS presents the sub-carrier spacing, SSB presents the single sideband, and BW is bandwidth. 

\begin{table}[h]
\centering
\caption{The parameters of 5G signals}
\label{tab:datasetParameter}
\begin{tabular}{|l|c|c|}
\hline
\textbf{Channel Parameter}& {\textbf{Value }} & Unit \\
 \hline 
5G BW & [10 15 20 25 30 40 50 ] & MHz\\
 \hline 
 5G SCS & [15 30 ] & kHz\\
 \hline 
 5G SSB Block Pattern & ["Case A" "Case B'] & - \\
 \hline 
  5G Period & [20] & ms\\
 \hline 
  LTE Reference Channel & ["R.2", "R.6", "R.8", "R.9"] & - \\
 \hline 
  LTE BW & [10 5 15 20 ] & MHz\\
 \hline 
  LTE Dublex Mode & FDD & -\\
 \hline 
  Channel SNR & [40 50 100] & dB\\
  \hline 
  Channel Doppler & [0 10 500] & Hz\\
  
 \hline 
\end{tabular}
\label{tab:dataset}
\end{table}

\begin{figure}[htbp]
    \centering
    \includegraphics[width=\linewidth]{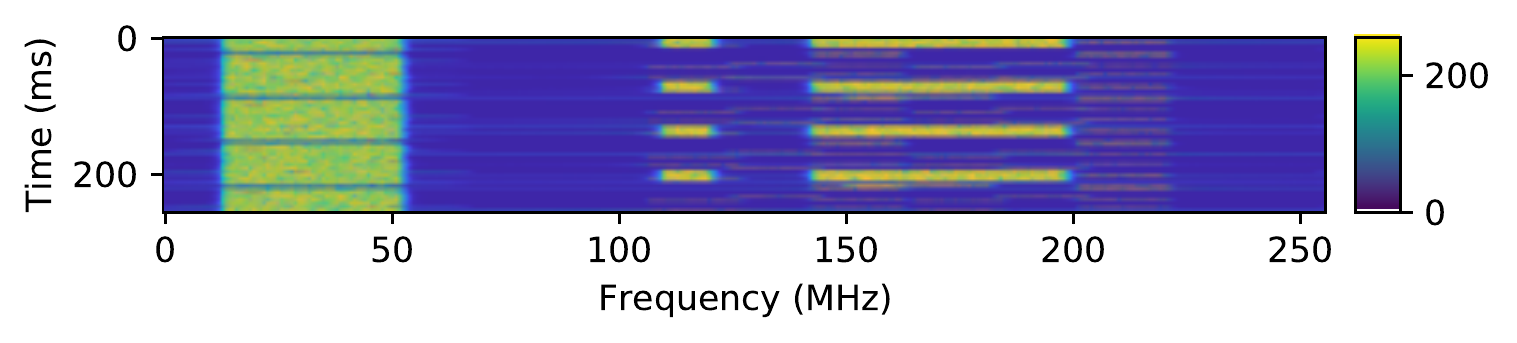}
    \caption{Example of 5G NR signals in the spectrogram}
    \label{fig:example_dataset_instance}
\end{figure}

Figure \ref{fig:example_dataset_instance} shows one of the example instances from the dataset.

\section{Experiment Results}\label{sec:experiments}

\begin{figure}
    \centering
    \includegraphics[width=0.9\linewidth]{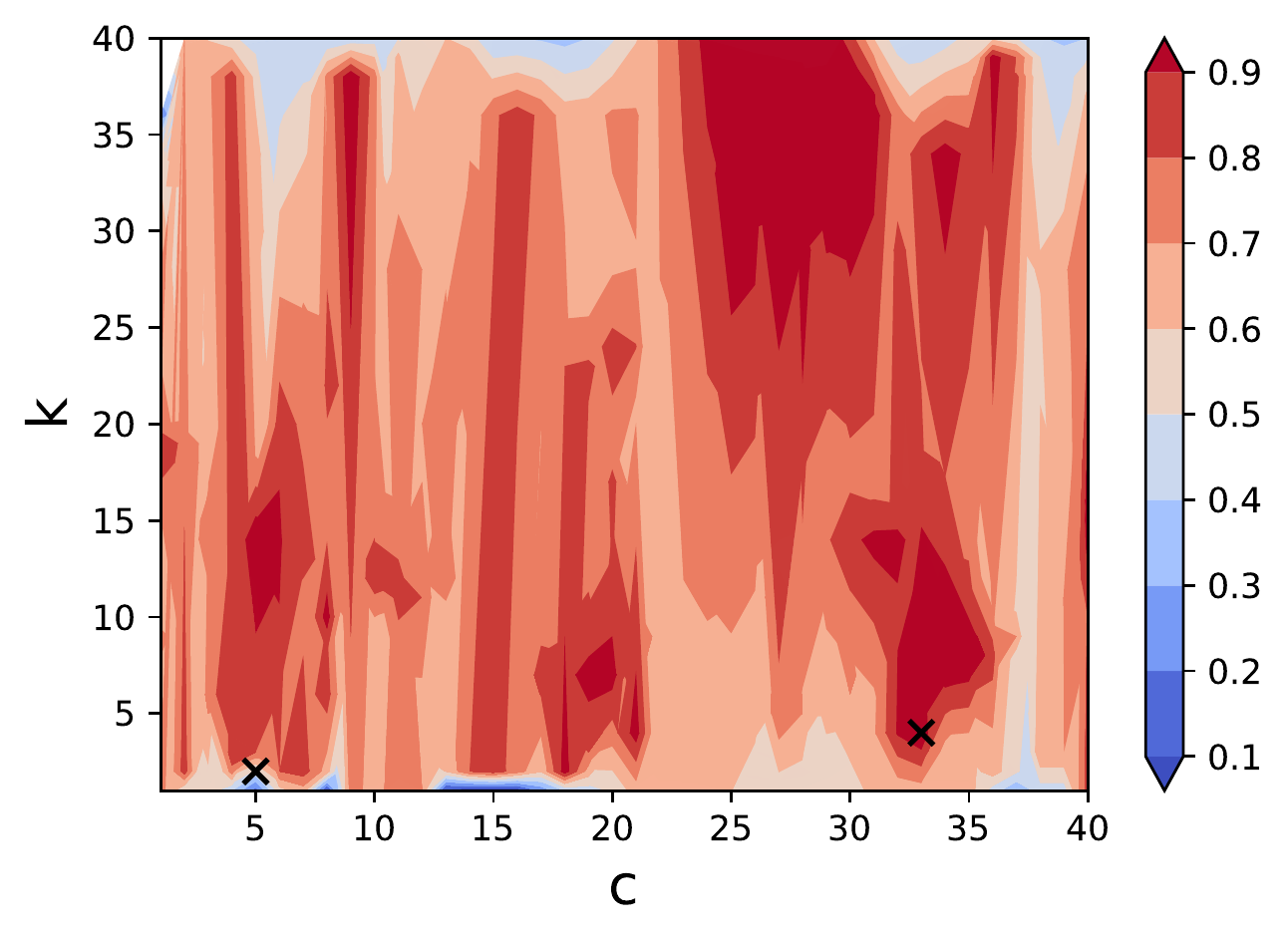}
    \caption{Heatmap plot of entropy values with different $c$ and $k$ values.}
    \label{fig:ent_heatmap}
\end{figure}

In this section, we evaluate the proposed 5G-SRNG algorithm to assess its performance. The results are visualized in Figure \ref{fig:ent_heatmap}, which presents the entropy values obtained with varying values of $c$ and $k$. The results demonstrate that the entropy values remain relatively constant with different values of $c$ and $k$. This suggests that the proposed 5G-SRNG algorithm can produce stable and secure random numbers. These findings are significant as they reinforce the reliability of the 5G-SRNG algorithm as a source of randomness for cryptographic applications.

According to the figure, the maximum entropy value is obtained at  $c = 33$ and $k = 4$, and the lowest entropy value was obtained at $c = 5$ and $k = 2$.

In this study, we evaluate the quality of random numbers generated by the 5G-SRNG algorithm. To assess the quality of the generated random numbers, we employ the NIST Statistical Test Suite, an open-source suite of statistical tests for randomness developed by the National Institute of Standards and Technology (NIST) in the United States \cite{10.5555/2206233}. The NIST Statistical Test Suite is a well-established tool for testing the quality of random numbers and provides a comprehensive evaluation of the randomness of the generated numbers.

The results of the 5G-SRNG using the NIST Statistical Test Suite are presented in Table \ref{tb:results} with four different combinations of $c$ and $k$ values: $10 \times 10$, $20 \times 20$, $30 \times 30$, and $40 \times 40$. The NIST Statistical Test Suite results indicate that the 5G-SRNG passes all statistical tests with a p-value greater than 0.01. This suggests that the 5G-SRNG can generate high-quality random numbers that exhibit desirable properties of randomness.
\begin{table*}[!ht]
\begin{center}
\caption{NIST Results of 5G-SRNG}
\label{tb:results}
\begin{tabular}{llrlrlrlr||cc}
\hline \rowcolor[HTML]{CCCCCC} 
                                Tests &  \multicolumn{2}{c}{$10 \times 10$} &  \multicolumn{2}{c}{$20 \times 20$} &  \multicolumn{2}{c}{$30 \times 30$} &  \multicolumn{2}{c}{$40 \times 40$} &  Adam et al. \cite{9190421} & Gao et al. \cite{9675830}\\
\hline \hline
                            Frequency &       \checkmark &   0.388820 &       \checkmark &   0.458808 &       \checkmark &   0.155274 &       \checkmark &   0.570750 & 0.8198 & 0.5790 \\
                      Block Frequency &       \checkmark &   0.060699 &       \checkmark &   0.061344 &       \checkmark &   0.204953 &       \checkmark &   0.719350 & 0.0818 & 0.4865 \\
                                  Run &       \checkmark &   0.653569 &       \checkmark &   0.679101 &       \checkmark &   0.639239 &       \checkmark &   0.492282 & 0.6168 & 0.5521 \\
            Run (Longest Run of Ones) &       \checkmark &   0.983331 &       \checkmark &   0.681757 &       \checkmark &   0.808265 &       \checkmark &   0.570061 &0.57581 & 0.5701\\
                   Binary Matrix Rank &       \checkmark &   0.080511 &       \checkmark &   0.053271 &       \checkmark &   0.615934 &       \checkmark &   0.792660 & 0.4942 & - \\
Discrete Fourier Transform (Spectral) &       \checkmark &   0.781415 &       \checkmark &   0.677254 &       \checkmark &   0.298093 &       \checkmark &   0.677254 & 0.1695 & 0.5047\\
    Non-overlapping Template Matching &       \checkmark &   0.762402 &       \checkmark &   0.518053 &       \checkmark &   0.818179 &       \checkmark &   0.067297 & 0.09245& 0.5028\\
        Overlapping Template Matching &       \checkmark &   0.874267 &       \checkmark &   0.368007 &       \checkmark &   0.611259 &       \checkmark &   0.127222 & 0.1564 & 0.5107\\
                Universal Statistical &      \xmark &  -1.000000 &      \xmark &  -1.000000 &      \xmark &  -1.000000 &      \xmark &  -1.000000 & 0.3457 & 0.4600\\
                    Linear Complexity &       \checkmark &   0.460431 &       \checkmark &   0.248459 &       \checkmark &   0.160908 &       \checkmark &   0.158490 & 0.0378 & 0.4898\\
                               Serial &       \checkmark &   0.311347 &       \checkmark &   0.923500 &       \checkmark &   0.379838 &       \checkmark &   0.601499 &- & 0.5664 \\
                  Approximate Entropy &       \checkmark &   0.145598 &       \checkmark &   0.109796 &       \checkmark &   0.236354 &       \checkmark &   0.232067 & 0.9999 & 0.4388\\
            Cumulative Sums (Forward) &       \checkmark &   0.557047 &       \checkmark &   0.319382 &       \checkmark &   0.280911 &       \checkmark &   0.110659 & 0.8389 & 0.5237 \\
           Cumulative Sums (Backward) &       \checkmark &   0.369108 &       \checkmark &   0.836898 &       \checkmark &   0.203432 &       \checkmark &   0.354357 & 0.66075 & -\\
\hline
\end{tabular}

\end{center}
\end{table*}

The results from the NIST Statistical Test Suite demonstrate that 5G-SRNG can generate high-quality random numbers, as evidenced by its successful passing of all statistical tests with a p-value greater than 0.01. The stability and security of the random numbers generated by 5G-SRNG were also maintained under varying values of $c$ and $k$. These results establish that 5G-SRNG can be utilized confidently to generate high-quality random numbers in 5G networks. From comparing the result of authors in  \cite{9190421} and the result of our work, it appears that the result of our experiments for some of the tests are better. Gao et al. \cite{9675830} presents a unified design of Physically Unclonable Function (PUF) and TRNG using an entropy source chip based on random access memory (RRAM). The design can be used for authentication systems with an authentication error rate approaching 0\% in IoT security applications. The design was tested across a wide range of temperatures and supply voltages. We add their results to the table.

We further evaluate the performance of the proposed 5G-SRNG by assessing its latency. This is a critical metric for implementing random number generators, especially for cryptography applications. We measure the latency of 5G-SRNG by utilizing a timer, which records the duration required to generate random numbers using the 5G-SRNG. The timer results are presented in Table \ref{tb:latency}. Our results indicate that the latency of 5G-SRNG is exceptionally low, demonstrating its suitability for cryptography applications that require fast and efficient random number generation.

\begin{table}[!htbp]
\centering
\caption{Latency of 5G-SRNG}\label{tb:latency}
\begin{tabular}{|c|c|c|c|}
\hline \rowcolor[HTML]{CCCCCC} 
\textbf{n(bits)} & \textbf{Latency (ms)} \\ \hline \hline
4096 & 329 \\ \hline
8192 & 338  \\ \hline
16384 & 344 \\ \hline
32786 & 396 \\ \hline
\end{tabular}

\end{table}

\section{Conclusion and Future Work}\label{sec:conclusion}

This paper proposed a 5G Spectrogram-based Cryptographically Secure pseudo RNG (5G-SRNG) for key exchange protocols. The 5G Spectrogram is used as an entropy source for the 5G-SRNG. The 5G Spectrogram is used to generate high-quality random numbers. We use the NIST Statistical Test Suite to evaluate the quality of random numbers. The results show that the quality of random numbers is improved using the 5G Spectrogram. The NIST Statistical Test Suite shows that the random numbers generated with the 5G-SRNG have a low p-value, indicating that the random numbers are high-quality. The proposed 5G-SRNG can be used for cryptographic protocols to protect the communication between IoT devices.

In future work, we will use different entropy sources and evaluate the quality of random numbers generated by 5G-SRNG. We will also use 5G-SRNG in different cryptographic protocols and analyze the performance of cryptographic protocols.

\bibliography{refs}
\bibliographystyle{IEEEtran}

\end{document}